\documentclass{article}
\usepackage{spconf,amsmath,graphicx,url,subfig, cite,booktabs}
\usepackage{amsfonts,amssymb}
\usepackage{xcolor}
\usepackage{makecell}
\usepackage[hidelinks]{hyperref}
\usepackage{svg}
\title{{TaylorAECNet}: A TAYLOR STYLE NEURAL NETWORK FOR FULL-BAND ECHO CANCELLATION}

\name{
\begin{tabular}{c}
\it Weiming Xu$^{1,*}$\thanks{*Work performed during an internship at Elevoc.}, Zhihao Guo$^2$
\end{tabular}
}
\address{
  $^1$Audio, Speech and Language Processing Group (ASLP@NPU), \\ Northwestern Polytechnical University, Xi'an, China\\
  $^2$Elevoc, Shenzhen, China
  }

\begin{document}
\ninept
\maketitle
\vspace{-6pt}
\begin{abstract}
% AEC challenge
% AEC-CHallenge

This paper describes aecX team's entry to the ICASSP 2023 acoustic echo cancellation (AEC) challenge. Our system consists of an adaptive filter and a proposed full-band Taylor-style acoustic echo cancellation neural network (TaylorAECNet) as a post-filter. Specifically, we leverage the recent advances in  Taylor expansion based decoupling-style interpretable speech enhancement~\cite{li2022taylor} and explore its feasibility in the AEC task. Our TaylorAECNet based approach achieves an overall mean opinion score (MOS) of 4.241, a word accuracy (WAcc) ratio of 0.767, and ranks 5th in the non-personalized track (track 1).

\end{abstract}

\begin{keywords}
Acoustic echo cancellation, noise suppression, Taylor expansion
\end{keywords}

\vspace{-12pt}
\section{Introduction}
\vspace{-6pt}
\label{sec:intro}

The acoustic echo cancellation (AEC) challenge series has provided a common platform to benchmark modern AEC techniques. The fourth edition of this challenge held in ICASSP2023 particularly addresses full-band signals with general AEC track and personalized AEC track. We submit a hybrid system to the general AEC track, which integrates a DSP based adaptive filter with a neural network based post filter.

Recently, there has been a trend in interpretable speech enhancement by decoupling the difficult task into easier interpretable subtasks. Following this direction, TaylorSENet~\cite{li2022taylor} imitates the form of Taylor expansion to split the denoising task into a combination of zero-order derivatives (in the real-valued domain) and multiple higher-order derivatives (in the complex-valued domain), achieving promising denoising performance. Inspired by this work, we propose a full-band Taylor-style acoustic echo cancellation neural network \textit{TaylorAECNet} as a post-filter cascaded with an adaptive filter to solve the full-band echo cancellation task. To reduce complexity, we design the Taylor-style decoupling network with only zero-order and first-order modules and made the following modifications to better fit the full-band residual echo removal task.
% Based on Taylor-style decoupling network with zero-order and first-order modules, we further made the following modifications to better fit the full-band residual echo removal task.
\begin{itemize}
        \item In order to utilize phase information for the zero-order module, a gated version of the phase encoder~\cite{zhang2022multi} is used instead of modulo operation to get the magnitude feature as zero-order module input.
	\item Pseudo quadrature mirror filter bank (PQMF) is used to reduce the complexity caused by full-band signal processing.
	\item Temporal and frequency convolution module (TFCM)~\cite{zhang2022multi} is introduced to improve the receptive field. Similar to~\cite{zzhang2022multi}, auxiliary voice activity detection (VAD) task is added and echo weighted loss is introduced.
 % 回声抑制量，近端更友好
\end{itemize}
According to the challenge results, our system ranked 5th in the general AEC track.

\vspace{-12pt}
\section{Proposed Method}
\vspace{-4pt}
\label{sec:format}
The AEC task can be described as:
\vspace{-4pt}
\begin{equation}
    d(n) = s(n) + h(n)\ast x(n) + v(n).
\vspace{-3pt}
\end{equation}
The near-end microphone signal $d(n)$ is composed of the near-end speech $s(n)$, the background noise $v(n)$, and the echo signal $ h(n)\ast x(n)$. 
The echo signal is generated by far-end signal $x(n)$ propagated through echo path $h(n)$, where $n$ denotes the sample index. The AEC task is to cancel $h(n)\ast x(n)$ from $d(n)$ given $x(n)$.

\begin{figure}[t]
	\centering
	\includegraphics[scale=0.25]{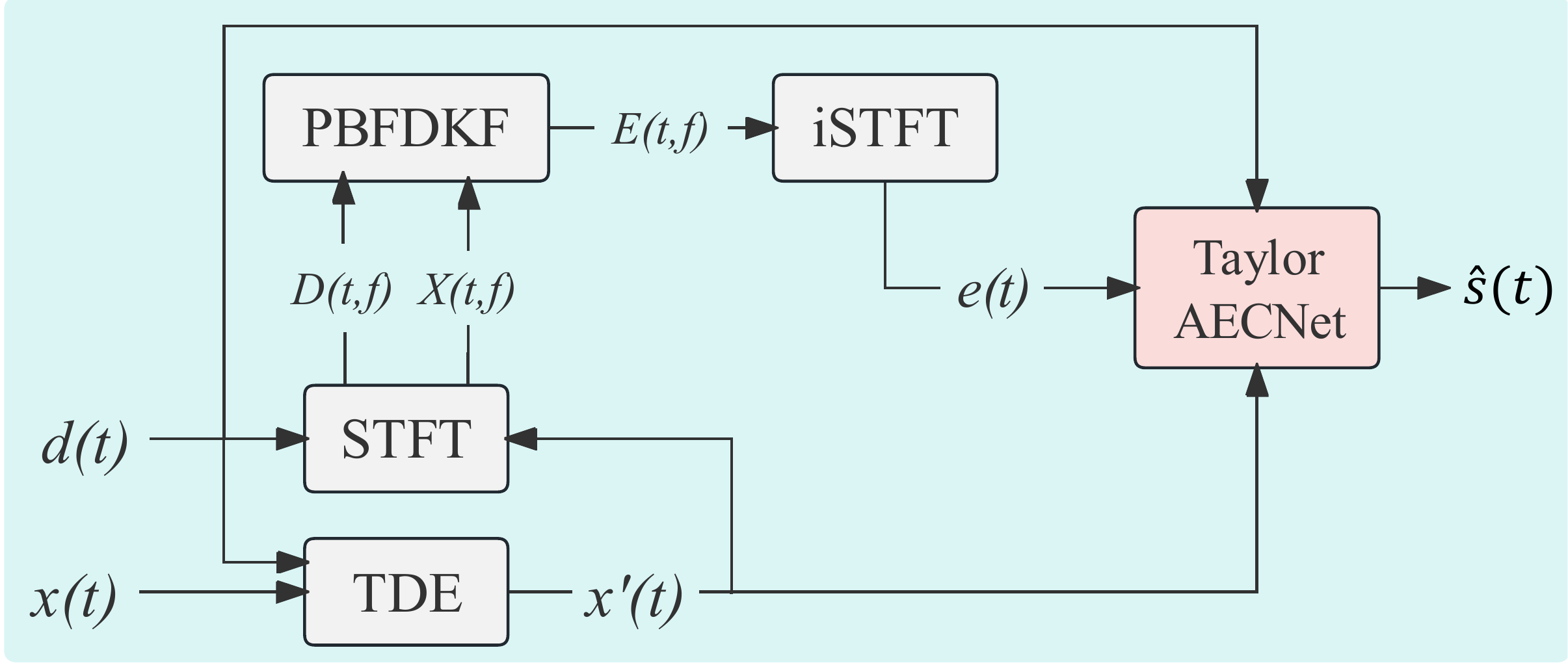}\vspace{-0.2cm}
	\caption{System diagram.}
	\label{fig:overview}  
\end{figure}

As shown in Fig.~\ref{fig:overview}, our system consists of a time delay estimation (TDE) module, an adaptive filter, and a full-band Taylor-style acoustic echo cancellation neural network (TaylorAECNet) as a post-filter. $d(n)$ and $x(n)$ are first aligned using GCC-PHAT as the TDE module to obtain the time-aligned reference signal $x'(n)$. The error signal $e(n)$ is generated using $x'(n)$ and $d(n)$ by an adaptive filter which is a partitioned-block-based frequency domain Kalman filter (PBFDKF). 
%$D, E$ and $X'$ are the frequency domain representation of $d, e$ and $x'$, respectively. 
Finally, $d(n)$, $e(n)$ and $x'(n)$ are stacked and fed to the TaylorAECNet post-filter.
% The TDE module uses the GCC-PHAT algorithm, which estimates the time delay between $d(n)$ and $x(n)$.% 
\vspace{-6pt}
\subsection{TaylorAECNet post-filter}
\vspace{-4pt}

\begin{figure}[h]
	\centering
	\includegraphics[scale=0.4]{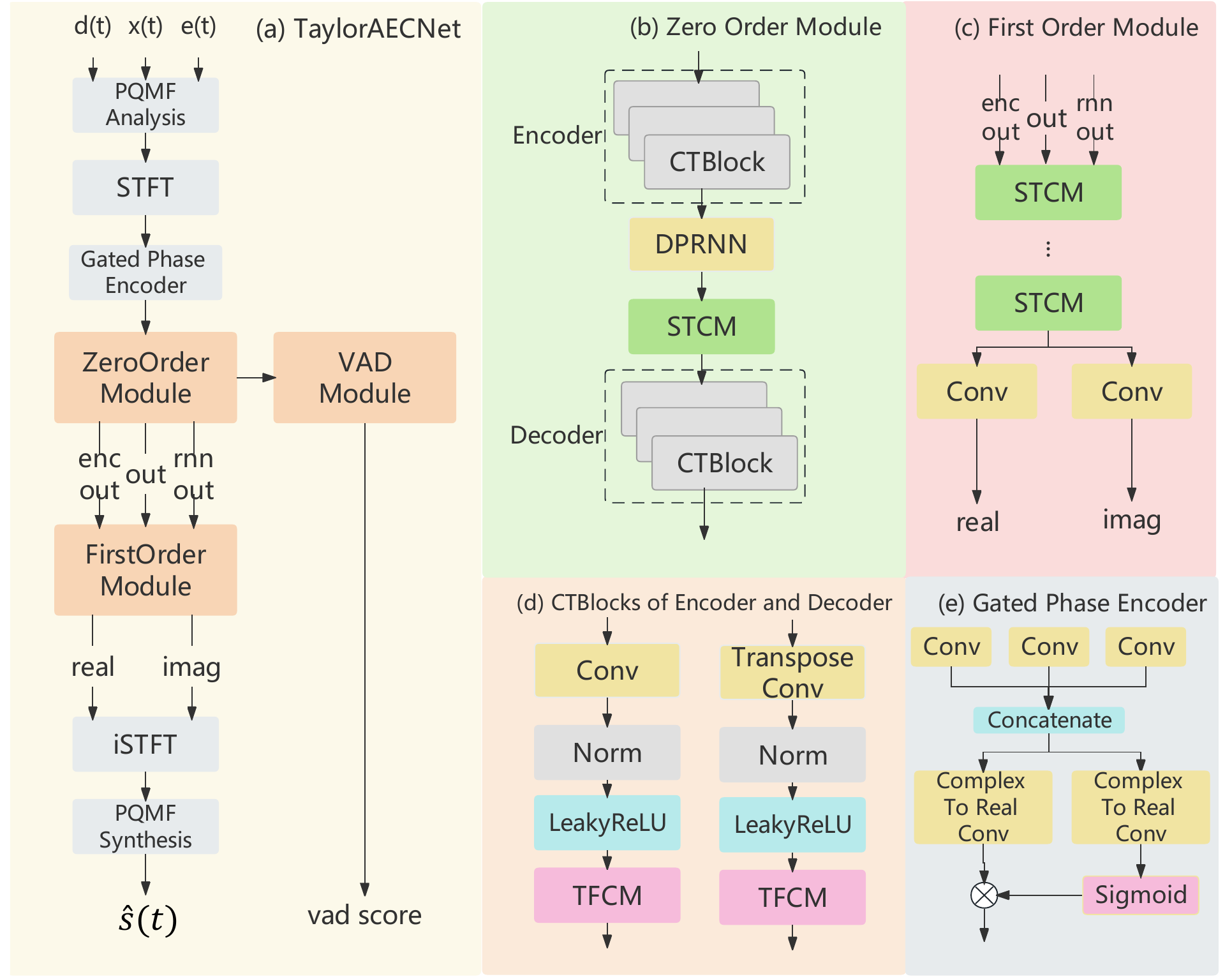}\vspace{-0.2cm}
	\caption{The detailed design of TaylorAECNet.}\vspace{-0.5cm}
	\label{fig:detail}
	\vspace{4pt}  
\end{figure}

TaylorAECNet consists of three modules,  zero-order module (ZOM), first-order module (FOM) and voice activity detection (VAD) module shown as Fig.~\ref{fig:detail}(a). 

As shown in Fig.~\ref{fig:detail}(b), ZOM is designed as a classical UNet-style structure and it accepts the magnitude feature from the gated phase encoder. The encoder is a stack of Convolution-TFCM Block (CTBlocks), and each CTBlock contains a convolution layer followed by a batch normalization layer, leakyReLU layer and TFCM~\cite{zhang2022multi}. The decoder has the same structure as the encoder, but the convolution layer is replaced by the transpose convolution layer. We use dual path RNN (DPRNN) and squeezed temporal convolution module (STCM)~\cite{li2021two} to further improve ZOM performance.

Fig.~\ref{fig:detail}(c) shows the detail of FOM which consists of several STCM layers and two convolution layers -- a complex-to-real convolution layer and a complex-to-imaginary convolution layer.

The structure of the gated phase encoder is shown in Fig.~\ref{fig:detail}(e), which is an improved version of phase encoder\cite{zhang2022multi} with an additional gated convolution. Specifically, the gated phase encoder has three complex-valued convolution layers to receive $D(t,f)$, $E(t,f)$, and $X(t,f)$ respectively. The complex-to-real layer consists of a complex-valued convolution layer and a modulo operation. We use a gated phase encoder instead of directly performing modulo operations on $D(t,f)$, $E(t,f)$, and $X(t,f)$.

We introduce a voice activity detection (VAD) module to improve the echo suppression performance. The VAD module is the same as that in~\cite{zzhang2022multi}. By learning the VAD state of the microphone signal, the model will be more inclined to preserve near-end speech. 
% When near-end speaker is not activated, the model will improve noise and echo suppression. Otherwise, the model will be more inclined to preserve near-end speech and reduce distortion.

\vspace{-12pt}
\subsection{Loss function}
\vspace{-5pt}
We use echo weighted loss~\cite{zzhang2022multi} with an extra asymmetric loss $\mathcal{L}_{\text{asym}}$~\cite{braun2022task}. The final loss function is
\vspace{-6pt}
\begin{equation}
    \mathcal{L} = \mathcal{L}_{\text{echo-weighted}} + \mathcal{L}_{\text{asym}} + 0.2*\mathcal{L}_{\text{mask}} + 0.1*\mathcal{L}_{\text{vad}}.
\vspace{-4pt}
\end{equation}
The definition of ${L}_{\text{mask}}$, ${L}_{\text{echo-weighted}}$, and $\mathcal{L}_{\text{vad}}$ remains the same 
as~\cite{zzhang2022multi}.  $\mathcal{L}_{\text{echo-weighted}}$ makes the model pay more attention to the suppression of echoes by echo weighting but may cause distortion of the near-end speech. So, we add $\mathcal{L}_{\text{asym}}$ to alleviate the distortion problem and constrain the spectrum.

\vspace{-12pt}
\section{Experiments}
\vspace{-8pt}
\subsection{Dataset}
\vspace{-6pt}

We use the clean speech provided by the 5th deep noise suppression (DNS5) as the near-end signal and the reference signal. The noise set in DNS5 is used as the noise signal. For the echo signal, we use all the synthetic echo signals and real far-end single-talk recordings provided by the AEC challenge, which covers a variety of voice devices and echo signal delay. Furthermore, we also use speech in the DNS5 dataset to simulate echo data by convolving with 100,000 simulated room impulse responses (RIR) generated by the image method. 

The training set has 600 hours of data in total, consisting of 300 hours of data with simulated echo and 300 hours of data with real-recorded echo. The development and test sets share the same aforementioned generation method, which contains 10 and 5 hours of data, respectively. 

\vspace{-12pt}
\subsection{Experimental setup}
\vspace{-4pt}
The proposed model uses a 20ms window length with a 10ms frame shift for STFT. Noam is adopted as the learning rate decreasing method described as
\vspace{-6pt}
\begin{equation}
    lr=d^{-0.5} \cdot \min(\text{step}^{-0.5}, \text{step} \cdot \text{warmup\_step}^{-1.5}) 
\vspace{-4pt}
\end{equation}
where $d=1e-3$ and warm\_up=5000. All the models are trained with Adam optimizer for 10 epochs. The PQMF splits the signal into 4 sub-band signals. Each TFCM has 6 layers. The STCM in ZOM has 1 layer and 128 hidden channels while the STCM in FOM has 2 layers and 64 hidden channels. 

\vspace{-12pt}
\subsection{Results and analysis}
\vspace{-6pt}
Table~\ref{tab:table1} shows the experimental results on the simulated test set. Ablation is studied on models trained using a subset with 100 hours of simulated data.
Here base-TaylorAECNet is a simplified version for ablation, which removes TFCM and gate PE parts. When adding TFCM to base-TaylorAECNet, all our metrics on the test set are improved. Further adding the gated phase encoder, which results in the complete TalyorAECNet, the metrics for ST-FE and DT are improved, but the metric for ST-NE is slightly decreased. 

We finally train the complete TalyorAECNet using the entire 600 hours of data and the metrics on the simulated test set are shown in the last row of Table \ref{tab:table1}. We use this model to process the blind test clips and the challenge results are shown in Table \ref{tab:table2}. Our proposed TaylorAECNet approach surpasses the baseline by a large margin in both Overall MOS and WAcc. The Final Score of our system is 0.803 which outperforms the baseline with a 0.067 gain, leading to the 5th in the general AEC track. We perform ONNX implementation to the post filter for speed-up, the RTF of our proposed system is 0.224 tested on Intel(R) Xeon(R) Gold 5218 CPU @ 2.30GHz and the number of its parameters is 19.18M. 

\begin{table}[h]
\caption{Ablation on the simulated test set. }
\centering
\scriptsize

\resizebox{\linewidth}{!}{
\begin{tabular}{lccccc}
\toprule
  Model  & Para. (M) & \makecell[c]{ST-FE\\(ERLE)} & \makecell[c]{ST-NE\\(WB-PESQ)} & \makecell[c]{DT\\(WB-PESQ)} &  Data                \\  \midrule
Noisy     &   &   0    &  2.62     &  1.95   &                      \\
base-TaylorAECNet & 15.2 & 51.43    &   3.08    &  2.68  &                      \\
~~+TFCM        & 19.1 & 62.84    &   3.05    &  2.69   &  100h                   \\
~~~~+gated phase encoder  & 19.2  &  63.33  &  3.01  &  2.72  &                     \\  \midrule
TaylorAECNet     & 19.2  &  \textbf{66.39}    &  \textbf{3.09}    &  \textbf{2.81}  & 600h                  \\   
\bottomrule
\end{tabular}
}
\label{tab:table1}
\end{table}
\vspace{-12pt}

\begin{table}[h] 
    \small
    \caption{ System performance on the blind test set of the challenge.}
	\centering
    \setlength{\tabcolsep}{20pt}
	\resizebox{\linewidth}{!}{
	\begin{tabular}{lccc}
		\toprule
		Model    & \multicolumn{1}{l}{Overall MOS} & \multicolumn{1}{l}{WAcc} & \multicolumn{1}{l}{Final Score} \\  \midrule
		baseline & 4.013                           & 0.649                    & 0.736                           \\
		TaylorAECNet   & 4.241                           & 0.767                    & 0.803                           \\ \bottomrule 
	\end{tabular}
	}
	\label{tab:table2}
\end{table}

\vspace{-22pt}
\section{Conclusions}
\vspace{-4pt}
\label{sec:typestyle}

In this paper, we introduce our system for ICASSP 2023 AEC Challenge. We combine a partitioned-block-based frequency domain Kalman filter (PBFDKF) with a proposed TaylorAECNet to suppress echo and noise. Gated phase encoder and TFCM are used for better latent feature extraction. A VAD module and echo-weighted loss are introduced to suppress residual echo and preserve near-end speech quality as well. According to the final results, the proposed system ranked 5th in the general AEC track.
% 总结要把自己的工作重新整理一下，带上实验结论

\vspace{-6pt}
\scriptsize
\bibliographystyle{IEEEbib}
% \vspace{-6pt}
\bibliography{strings}
\end{document}